# Vertically Graded Anisotropy in Co/Pd Multilayers

B. J. Kirby,[1,*] J. E. Davies,[2,3,†] Kai Liu,[4] S. M. Watson,[1] G. T. Zimanyi,[4]

R. D. Shull,[2] P. A. Kienzle,[1] and J. A. Borchers[1]

[1]*Center for Neutron Research, NIST, Gaithersburg, MD  20899*

[2]*Metallurgy Division, NIST, Gaithersburg, MD  20899*

[3]*Advanced Technology Group, NVE Corporation, Eden Prairie MN  55344*

[4]*Physics Department, University of California, Davis, CA  95616*

**Abstract**

Depth-grading of magnetic anisotropy in perpendicular magnetic media has been predicted to reduce the field required to write data without sacrificing thermal stability.  To study this prediction, we have produced Co/Pd multilayers with depth-dependent Co layer thickness.  Polarized neutron reflectometry shows that the thickness grading results in a corresponding magnetic anisotropy gradient.  Magnetometry reveals that the anisotropy gradient promotes domain nucleation upon magnetization reversal - a clear experimental demonstration of the effectiveness of graded anisotropy for reducing write-field.

*Electronic mail:  brian.kirby@nist.gov

† Electronic mail: jdavies@nve.com

Ongoing demand to increase the storage density of magnetic recording media continues to drive development of advanced new magnetic nanostructures. The central issue is to develop a media that exhibits excellent thermal stability, while maintaining a low enough coercivity to facilitate practical writing of data. Several options currently under exploration rely on modifying the recording system as a whole, such as patterned media, and heat and microwave-assisted recording. Another approach is to modify the film microstructure, for example by tuning growth properties to promote perpendicular anisotropy in multilayer thin films (e.g., Co/Pt or Co/Pd),[1,2,3] or by using a combination of discrete low and high anisotropy layers to produce exchange-coupled composite (ECC) media.[4,5,6] Suess *et al.* have proposed that continuously varying the anisotropy along the length of a columnar grain could lead to an even better "graded" media, where the coercivity is further reduced due to the exchange coupling across the magnetically hard and soft layers while thermal stability is anchored by the domain wall energy in the hard layer.[7,8,9,10] Experimentally, however, it has been very challenging to directly resolve the anisotropy gradient. This is due to the difficulty in probing depth-dependent magnetic configurations, as well as the presence of interlayer exchange coupling that can mask signatures of an anisotropy gradient in standard magnetometry measurements. In this work we report a first direct measurement of a depth-dependent magnetic anisotropy gradient in Co/Pd multilayers where the Co thickness is varied throughout the film stack in order to grade the perpendicular anisotropy. Polarized neutron reflectometry (PNR), superconducting quantum interference device (SQUID) magneotmetry, and vibrating sample magneotmetry (VSM) measurements conclusively show that the Co thickness gradation results in a corresponding gradation in magnetic anisotropy, and that this anisotropy gradient facilitates domain nucleation upon magnetization reversal.

The samples were grown using room temperature DC magnetron sputtering, and consist of multilayer stacks of 30 layers of Co, each separated by 0.9 nm of Pd. The stacks were deposited on a 20 nm Pd seed layer on a Si substrate with a native oxide layer, and were capped with a 5 nm Pd layer. The first 15 Co/Pd bilayer repeats for each sample were nominally identical high anisotropy regions with Co thickness $t_{Co}$ = 0.3 nm. For the subsequent 15 Co/Pd bilayers, $t_{Co}$ varied differently for each of the four samples studied:

1) 15 repeats of $t_{Co}$ = 0.3 nm.
2) 15 repeats of $t_{Co}$ = 0.7 nm.
3) 8 repeats of $t_{Co}$ = 0.5 nm, and 7 repeats of $t_{Co}$ = 0.7 nm.
4) 15 Co bilayers where $t_{Co}$ progressively increased from layer to layer, from 0.3 nm to 1.1 nm.

For convenience, we refer to the samples in terms of one "layer" as encompassing a sub-stack of like $t_{Co}$ - thus, we refer to sample 1) as "monolayer", 2) as "bilayer", 3) as "trilayer, and 4) as "graded".

While $t_{Co}$ grading was expected to result in a corresponding anisotropy gradient, it was conceivable that such an effect could be masked by a dominating exchange coupling among differing $t_{Co}$ regions. To determine whether or not $t_{Co}$ grading actually leads to a graded anisotropy, the samples were studied with specular PNR - a technique that is sensitive to structural and magnetic depth profiles of thin films and multilayers.[11, 12] In particular, the technique is sensitive only to the in-plane component of the magnetization $M$, and is totally insensitive to magnetization normal to the sample surface.[12, 13] By performing PNR measurements as a function of increasing in-plane applied field ($H$), we were able to determine how spins at different depths of the multilayer stack responded to being pulled away from the perpendicular easy axis direction - thus enabling characterization of the depth-dependent anisotropy. PNR measurements were conducted using the NG-1 Reflectometer[14] at the NIST Center for Neutron Research. An incident monochromatic neutron beam was polarized to be alternately spin-up or spin-down relative to $H$. The non spin-flip reflectivity (with incident beam spin-up or spin-down), and the spin-flip reflectivity (up to down and down to up) were measured as a function of scattering vector $Q$. The samples' depth-dependent nuclear scattering length density $\rho_N(z)$ profiles (functions of the scattering potential of the constituent nuclei at different depth $z$ beneath the sample surface) and $M(z)$ profiles were determined by model fitting the PNR data, using the GA_REFL software package.[15] Measurements were conducted at room temperature under increasing $H$, starting out from an out-of–plane AC demagnetized state.

Representative fitted PNR spectra for the graded sample are shown in Figure 1. At $\mu_0 H$ = 6 mT (a), there is very little difference between the spin-up and spin-down reflectivities, and significant spin-flip scattering is observed, indicating an in-plane magnetization that is not collinear with $H$. As the field is increased to 0.66 T (b and c), the spin-flip scattering disappears, and the spin-up and spin-down reflectivities become progressively more split, indicating an increase in magnetization along the direction of $H$. The $\rho_N(z)$ and $M(z)$ depth profiles for the graded sample determined by fitting of the data are shown in Fig. 2. For simplicity, the PNR data are not modeled in terms of the very thin individual layers, but instead in terms of Co/Pd regions with like $t_{Co}$.[16] A Gaussian transition function of fitted width between the two layers is used in the model to account for the gradation in $M$ and nuclear composition. Given that the maximum $Q$ for these measurements is well below that of any Bragg diffraction peaks corresponding to the repeat thicknesses within the stacks ($Q$ = 2.9 nm$^{-1}$), this simple choice of model is completely valid. Panel a) of Fig. 2 shows the nuclear profile used to fit the data and clearly indicates the positions of the Pd cap, Co/Pd film, Pd seed layer, native oxide layer, and the Si substrate at increasing $z$. Notably, the model fitting is sensitive to a decrease in $\rho_N$ for the Co/Pd film near the sample surface. Since Co is a weaker nuclear scatterer than Pd,[17, 18] this is indicative of the increase in Co content as the Co layers become progressively thicker. Panel b) shows the magnitude of the depth-dependent, in-plane magnetization projection $M(z)$ as a function of $H$. For $\mu_0 H = 6$ mT and 100 mT, spin-flip scattering indicates that the in-plane magnetization was oriented away from $H$ at angles of 260º and 14º respectively. For all other fields, no spin-flip scattering was detected, indicating that the in-plane magnetization was collinear with $H$. The graded $t_{Co}$ region of the sample exhibits a much larger magnetization at all fields, but this is partially due simply to the increase in Co content corresponding to thicker Co layers. In order to distinguish the contributions of depth-variations in total moment and depth-variations in anisotropy to the observed magnetization gradient, panel c) shows each of the curves in panel b) normalized by the respective maximum values. If the individual Co layers all exhibit the same anisotropy (e.g., as a result of a completely dominant interlayer exchange coupling), the magnetizations of those layers should all respond to $H$ at the same rate, making the normalized curves identical. However, this is not the case, as the normalized profiles are strikingly different. This demonstrates that the sample truly does exhibit graded anisotropy, where anisotropy decreases with increasing $t_{Co}$. We note that models in which the magnetizations are *constrained* to be proportional produce significantly worse fits to the data, confirming our sensitivity to this result (see EPAPS supplementary document, E1). Models for the monolayer, bilayer, and trilayer data (see EPAPS, E2) yield results consistent with those of the graded sample. The monolayer is modeled well as a single uniform stack, and the bilayer and trilayer are modeled well with two and three layers, respectively. The bilayer and trilayer data are also consistent with perpendicular anisotropy that decreases with increasing $t_{Co}$.

With graded anisotropy confirmed to correspond with $t_{Co}$ grading, we now consider the results of room temperature magnetometry measurements (key parameters are summarized in Table 1). For $H$ along the perpendicular-to-plane easy axis, measurements were conducted with VSM. The major hysteresis loops for all four samples are wider near saturation and narrower near the coercive field $H_C$, as shown in Fig. 3a. These loops are characteristic of a three stage reversal process of: 1) irreversible domain nucleation and propagation, 2) reversible domain wall motion and 3) irreversible domain annihilation.[19] To obtain a more quantitative understanding of the switching, we performed a first order reversal curve (FORC) analysis of the magnetometry data - a technique that quantifies the irreversible components of magnetization in terms of a FORC distribution, $-\partial^2 M(H, H_R) / 2\partial H \partial H_R$, where $H_R$ is the reversal field.[20, 19, 21] FORC analysis is ideally suited for "fingerprinting" the presence of magnetic inhomogeneities, which are manifested as unique patterns in FORC diagrams. FORC distributions measured for our samples (EPAPS, Fig. E3) exhibit a horizontal ridge, a planar region, and a negative/positive pair of peaks - three features that correspond one-to-one with the three stages of reversal described above.[19] These features are summarized by integrating the FORC distribution over $H$, yielding the switching field distribution (FORC-SFD) shown in Fig. 3b. As the FORC-SFD is very sensitive to the onsets and endpoints of irreversible magnetization switching,[22, 23] we can confidently define the initial switching field (where $dM/dH_R$ becomes non-zero) as the nucleation field $H_N$, and define the point at which the FORC-SFD returns to $dM/dH_R = 0$ as the true saturation field ($H_S$). Although the coercive field $H_C$ served as the measure of switching and saturation fields in earlier theoretical work that considered isolated columnar grains,[7, 8, 9] here, $H_N$ and $H_S$ are more appropriate parameters for discussing the effects of graded anisotropy as for multilayer films strong inter-granular exchange coupling and dipole fields in the multi-domain state convolute the intrinsic $H_C$.

This effect on the hysteresis loops due to the thin-film geometry can be understood in terms of the magnetostatic energy of the uniformly magnetized film, or the demagnetization energy ($E_D = 2\pi M_S^2 t$, where $t$ is the total Co/Pd film thickness). Table 1 shows that the monolayer sample is the most stable in the uniformly magnetized state, while $E_D$ values for the other three samples are similar. This variation in demagnetization energy complicates direct, quantitative comparison of some parameters of interest. For example, if considering samples consisting of isolated grains, along the decreasing-field sweep the monolayer would be expected to exhibit the most negative saturation field.[7] However, the low $E_D$ of the monolayer sample results in it having the *least* negative saturation field. Thus, in this work, we focus on more qualitative, definitive trends among the samples.

Inspection of the descending-field branches of the easy-axis loops and the FORC-SFD (Figs. 3a and 3b, respectively) reveals that the graded sample nucleates at a higher field ($\mu_0 H_N = 0.16$ T) than the bilayer ($\mu_0 H_N = 0.14$ T), and monolayer ($\mu_0 H_N = 0.00$ T) samples. It is seemingly out of place that the trilayer sample ($\mu_0 H_N = 0.11$ T) does not exhibit $H_N$ intermediate between that of the bilayer and graded samples. This non-monotonic variation in $H_N$ is evidence that, as predicted,[7] the collective anisotropy of a given multi-anisotropy sample is extremely sensitive to the individual anisotropies and magnetizations of the constituent layers. Thus, the magnetic properties of this particular trilayer are not perfectly well tuned so as to achieve an improvement in $H_N$ over the bilayer. However, it is important that the general trend in $H_N$ clearly demonstrates that anisotropy grading facilitates magnetization reversal by promoting domain nucleation. This result constitutes a qualitative realization of the theoretically predicted[7, 8, 9, 10] reduction in write-field for graded anisotropy media.

In large part, the increased $H_N$ for the multi-anisotropy samples can be attributed to exchange coupling between hard and soft layers. Further evidence of this coupling is found in several other aspects of the magnetometry and PNR results. First, like the monolayer, the multi-anisotropy samples exhibit only one nucleation "step" with a precipitous magnetization drop (Figs. 3a and 3b). This demonstrates that when domain nucleation occurs, interlayer exchange coupling causes it do so throughout the entire thickness of the film, as opposed to soft layers nucleating at different fields than hard layers. Second, exchange coupling is manifested in the field-dependent in-plane magnetization of the hard $t_{Co} = 0.3$ nm regions common to all samples, as determined by PNR (see EPAPS, Figs. E8e, h, and k for full profiles). As shown in Fig. 4, the monolayer sample ($t_{Co} = 0.3$ nm) exhibits a smaller in-plane magnetization at $\mu_0 H \approx 650$ mT, than do the nominally identical $t_{Co} = 0.3$ nm regions in the bilayer, trilayer, and graded samples – indicative of the coupling between hard and soft regions. Third, along the hard in-plane axis (Fig. 3c, measured with SQUID), the bilayer, trilayer, and graded samples exhibit comparable saturation fields ($\mu_0 H_K \approx 2.3$ T) that are significantly smaller than that of the monolayer ($\mu_0 H_K \approx 3.0$ T). This implies that the net apparent anisotropy of the monolayer is greater than that of the multi-anisotropy samples, as theoretically predicted.[7]

Of particular interest is the *mechanism* by which $M$ reversal occurs in the graded sample, which as predicted exhibits maximum improvements in nucleation and saturation fields. Simulations suggest that graded anisotropy columns should reverse $M$ via nucleation of a partial vertical domain wall.[7, 8, 9, 10] The graded sample (open triangles in Fig. 3a) exhibits a perpendicular hysteresis loop remarkably consistent with this prediction. As $H$ is reduced from saturation, the graded sample shows a distinctive gradual decrease in magnetization at $\mu_0 H = 0.3$ T. As indicated by the FORC-SFD (Fig. 3b), this decrease in magnetization is reversible, as the onset of irreversible switching occurs at $\mu_0 H_N = 0.16$ T. The magnetization depth profile of the graded sample (Fig. 2b) reveals a large low field magnetization for the soft near-surface layers, and shows that those layers begin to saturate in-plane at $\mu_0 H \leq 0.35$ T. Thus, we conclude that as $H$ is reduced from magnetic saturation along the perpendicular easy axis, the softest layers relax *into the plane*. This indicates that for the graded sample, the magnetostatic energy is indeed minimized during reversal through introduction of a partial vertical domain wall.

In conclusion, we have directly observed a vertically graded magnetic anisotropy in thickness-graded Co/Pd multilayer films. Neutron scattering measurements reveal a graded anisotropy profile where anisotropy is reduced in regions with thicker Co layers. Magnetometry results show that along a decreasing-field sweep, anisotropy grading facilitates domain nucleation – a clear demonstration of enhanced writeability theoretically predicted for graded media.[7, 8, 9, 10] Exchange coupling between regions of differing anisotropy is shown to play an important role in the magnetization reversal of graded anisotropy samples, by reducing the net anisotropy and easing the reversal of the hardest layers. Our results experimentally demonstrate the concept that graded media would have a distinct write field advantage compared to constant anisotropy and possibly exchange-coupled composite media.

Work at UCD has been supported in part by CITRIS and NSF-ECCS 0925626. J.E.D. and K. L. acknowledge support from a National Research Council postdoctoral fellowship, and a UCD Chancellor's Fellowship, respectively. The authors thank Peter Greene of UCD for technical assistance, and Brian Maranville of NIST for valuable discussions.

11[17] H. Glattli, G. L. Bacchella, M. Fourmond, A. Malinovski, P. Meriel, M. Pinot, P. Roubeau, and A. Abragam, J. Phys. **40**, 629 (1979).

[18] J. W. Cable and E. O. Wollan, Phys. Rev. **140**, 2003 (1965).

[19] J. E. Davies, O. Hellwig, E. E. Fullerton, G. Denbeaux, J. B. Kortright, and K. Liu, Phys. Rev. B **70**, 224434 (2004).

[20] C. R. Pike, A. P. Roberts, and K. L. Verosub, J. Appl. Phys. **85**, 6660 (1999).

[21] H. G. Katzgraber, F. Pazmandi, C. R. Pike, K. Liu, R. T. Scalettar, K. L. Verosub, and G. T. Zimanyi, Phys. Rev. Lett. **89**, 257202 (2002).

[22] J. E. Davies, O. Hellwig, E. E. Fullerton, J. S. Jiang, S. D. Bader, G. T. Zimanyi, and K. Liu, Appl. Phys. Lett. **86**, 262503 (2005).

[23] J. E. Davies, O. Hellwig, E. E. Fullerton, M. Winklhofer, R. D. Shull, and K. Liu, Appl. Phys. Lett. **95**, 022505 (2009).

Table 1: Characteristic magnetic properties of the samples (uncertainties correspond to ± 1 $\sigma$).

| sample | easy-axis saturation magnetization $M_S$ (kA/m)§ | hard-axis saturation field $\mu_0 H_K$ (T) | magnetic layer thickness $t$ (nm)* | nucleation field $\mu_0 H_N$ (mT)† | easy-axis saturation field $\mu_0 H_S$ (mT)† | demagnetization energy $E_D$ (mJ/m$^2$) |
|---|---|---|---|---|---|---|
| monolayer | 530 ± 34 | 3.0 ± 0.2 | 36 ± 1 | 0 ± 2 | -340 ± 2 | 6.3 ± 0.8 |
| bilayer | 693 ± 44 | 2.3 ± 0.2 | 39 ± 1 | 140 ± 2 | -400 ± 2 | 11.8 ± 1.5 |
| trilayer | 647 ± 41 | 2.3 ± 0.2 | 39 ± 1 | 110 ± 2 | -390 ± 2 | 10.2 ± 1.3 |
| graded | 657 ± 52 | 2.3 ± 0.2 | 41 ± 2 | 160 ± 2 | -430 ± 2 | 11.1 ± 1.9 |

§ Magnetization is calculated by dividing the magnetic moment determined from magnetometry by the sample area, and by the magnetic layer thickness determined from PNR.

* From PNR measurements

† From FORC measurements

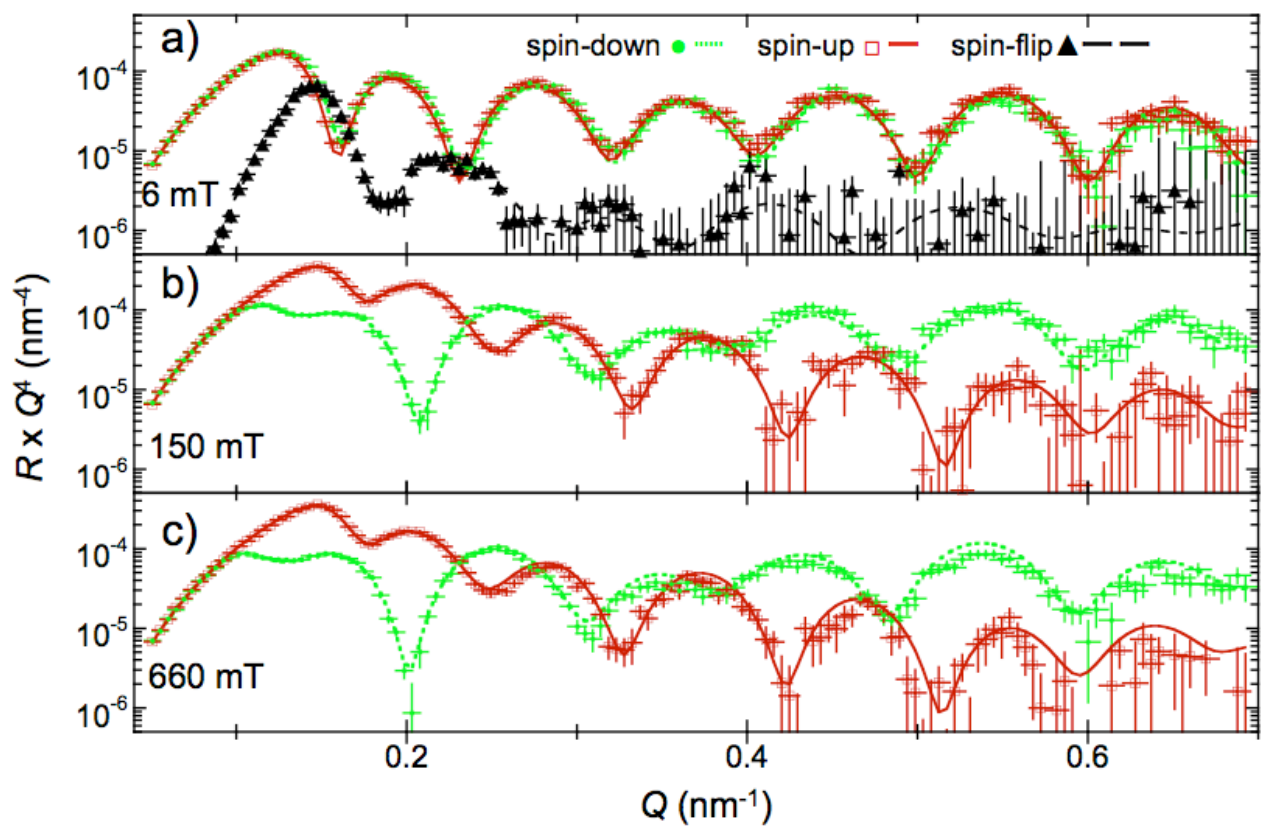

Figure 1 (color online): Fitted PNR spectra for the graded sample, measured at three different fields. Symbols are data, and lines are fits corresponding to the profiles in Figure 2. Error bars correspond to ± 1 $\sigma$.

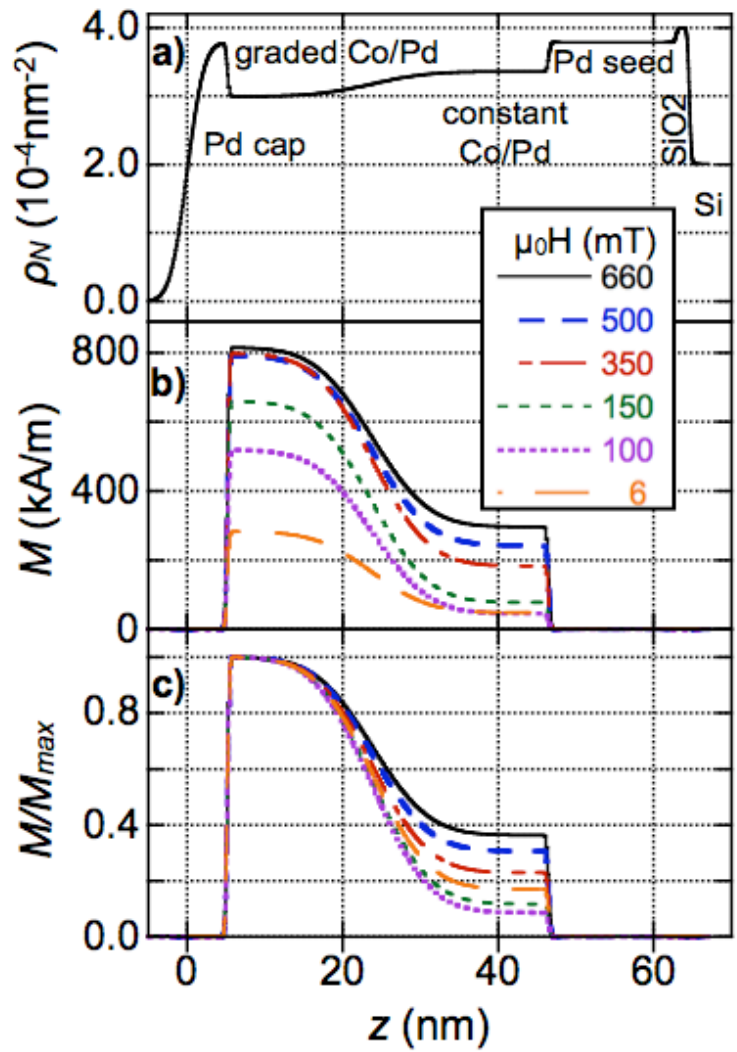

Figure 2 (color online): Models used to fit PNR spectra of the graded sample. a) shows the nuclear scattering length density, b) shows the field dependent magnetization profiles, and c) shows the magnetization profiles normalized by the respective maximum values. That the profiles in c) are different demonstrates that the sample exhibits graded anisotropy.

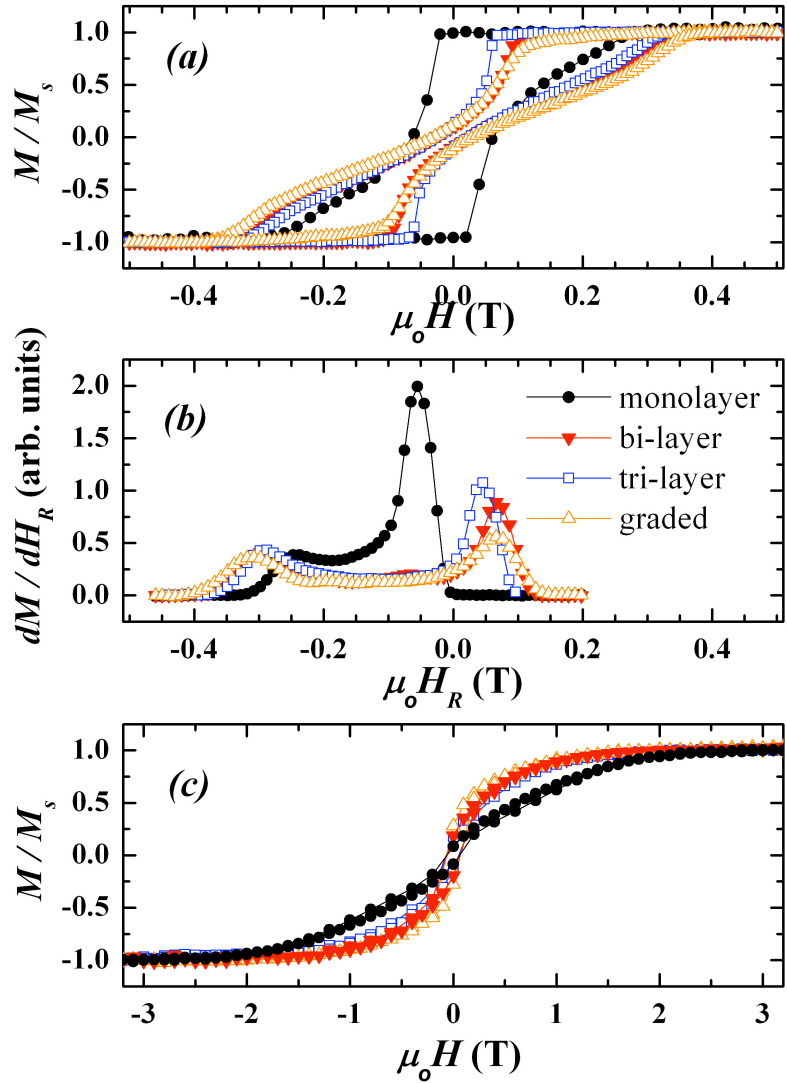

Figure 3 (color online): (a) Easy axis major loops measured with VSM, (b) FORC – switching field distributions and (c) hard axis major hysteresis loops measured with SQUID. The easy axis loops illustrate a clear reduction in nucleation field with anisotropy grading.

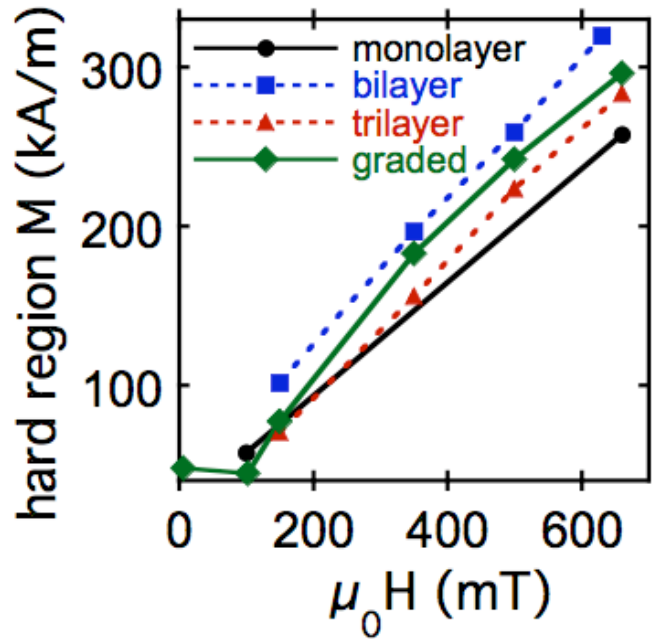

Figure 4 (color online): In-plane magnetizations of the "hard" $t_{Co} = 0.3$ nm regions of each sample as a function of in-plane applied field, as determined from PNR. The addition of "softer" layers promotes in-plane magnetization of the hard layers, a clear indication of exchange coupling.